\documentclass[english,aps]{revtex4-1}
\pdfoutput=1
\usepackage{amstext}
\usepackage{amssymb}
\usepackage{graphicx}
\begin{document}

\title{Carrier mobility modeling in highly inhomogeneous semiconductors}

\author{Algirdas Mekys$^{1,2}${*}, Vytautas Rumbauskas$^{2}$, Laurynas
Andrulionis$^{3}$ and Jurgis Storasta$^{2}$}

\affiliation{$^{1}$ Vilnius University, Institute of Theoretical Physics and
Astronomy, A. Gostauto St. 12, 01108 Vilnius, LITHUANIA }

\affiliation{$^{2}$ Vilnius University, Institute of Applied Research, Sauletekio
al. 9, III, LT-10222, Vilnius, LITHUANIA }

\affiliation{$^{3}$ Vilnius University, Faculty of Physics, Sauletekio al. 9,
III, LT-10222, Vilnius, LITHUANIA. }

\email{algirdas.mekys@ff.vu.lt}

\begin{abstract}
The large scale various shapes and orientation defects influence into
carrier scattering was theoretically analyzed using Monte Carlo method
and compared to the experimental measurements. It was shown how the
large scale defects screen themselves from the interaction with free
charge carriers. The comparison of quantum and classical transport
regimes was performed.
\end{abstract}


\maketitle

\section{Introduction}

One of the most important parameter describing semiconductor material
is the carrier mobility. It is directly related to the speed of electronic
devices made from that material. The mobility gives information about
the crystal quality of the material, amount of impurity or other defects.
The widely used mobility measurement methods like Hall, magnetoresistivity,
time-of-flight or others enables to find the mobility values mostly
in homogeneous materials, however they are often applied for the highly
inhomogeneous ones. The properties of such materials usually are difficult
to describe because they contain separate regions of the materials
with different properties, which do not manifest as simple average.
These materials are more similar to metamaterials, which distinguish
global from local characteristics. As an example of this manifestation
can be a conductive wire surrounded by electrical isolator. Thus the
mobility in the wire can be measured, but not in the isolator, which
may have high mobility but low density of charge carriers. In such
samples the electrical screening or short-cutting may disable mobility
calculations from Hall effect at all, like in \cite{key-1}, where
neutron irradiation introduced clusters of p-type point defects into
n-type crystals. However, this was not the case of carrier scattering
peculiarities. The carrier mobility in these experiments was extracted
from the simultaneously measured magnetoresistivity effect, showing
the irradiation unaffected crystal parts, which had high carrier mobility.
Very similar behavior of these two mobility measurements also were
observed in samples after the defect engineering \cite{key-2} or
self formation of oriented large scale defects \cite{key-3,key-4}
. In both cases the Hall voltage signal was anomalously low and the
magnetoresistivity effect was distinctly sensitive to the orientation
of the samples. The effect was not attributed to the change of the
carrier scattering. It was thought the effect rises from the measurement
particularity but not the carrier transport. Thus in the presence
of such kind large-scale defects the question was unanswered: when
the measurement results describing scattering processes can be distinguished
from the measurement peculiarities with partial incompatibility? To
answer this question, the modeling was performed with similar geometry
defects in the scope of quantum and classical transport mechanisms
with Monte Carlo method. The method was chosen because of implementation
simplicity and the availability of large computing resources. Several
types large scale defects were used for the carrier scattering investigations.
The supplemental research was performed as experimental modeling for
the certain large-scale defect configurations.

\section{The Theory Background}

\begin{figure}
\includegraphics{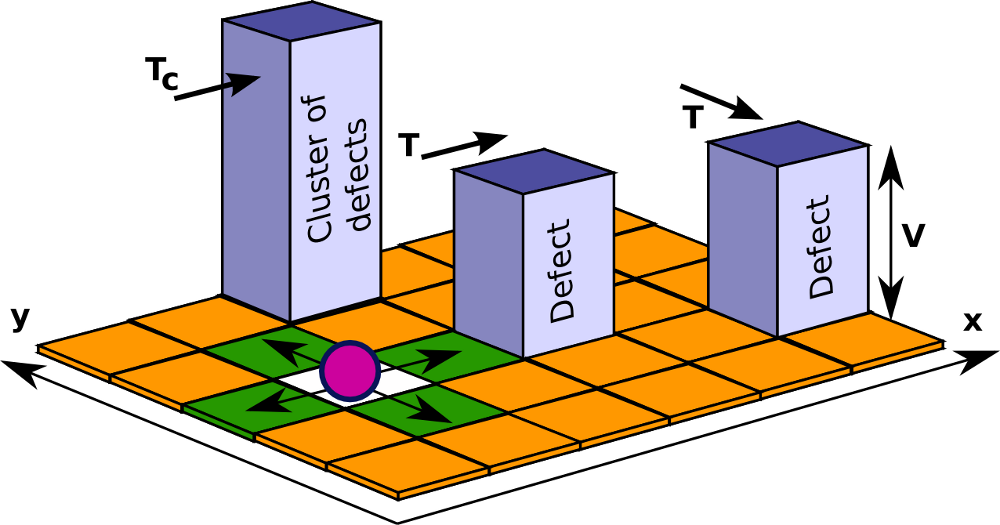}

\caption{Illustration of the modeling space (x,y) and parameters arrangement.
The single point defect occupies the space grid point with the potential
V, the cluster of defects occupies the same size in space but its
potential is higher. The transmission coefficient T depends on V and
is lower for the clusters comparing to the point defects. The particle
(circle) may jump into the nearest grid points in 4 directions with
the same probability except the cases when the probability is modified
with the presence of the defect. (Color online)}

\label{fig:Matrix}
\end{figure}

From the point of view of Monte Carlo method, the transport of carriers
in the crystal may be found by repeating the steps of a single carrier
walk and making the statistical average. The first assumption was
made that the carriers do not interfere between themselves. Then the
classical diffusion transport principles may be applied. In that case
the mobility of the particle is directly proportional to the diffusion
coefficient by Einstein relation $\left(\mu\sim D\right)$. The diffusion
of the particle is considered as random walk in the crystal and the
collision events occur with phonons and crystal structure irregularities.
The phonons at each time step appear randomly and in the modelling
they are considered as determining random direction for the next step.
Thus, the temperature is not included in this model other way and
its absolute value is not calculated. The mobility is reciprocally
proportional to the number of the scattering events and we will consider
the mobility only qualitatively. The screening of the internal potential
by free carriers is not included, so the model is limited to be suitable
for higher resistivity materials. The crystal structure defects may
be treated as particle scattering centres with the finite interaction
probability. In mesoscopic transport systems the particle transfer
through the defective region may be described using Landauer model
\cite{key-5}, which introduces the transfer coefficient T. The crystal
field distortion caused by defects may be considered as barriers or
wells in the paths of the quasi-free charge carriers. For the simplicity,
in our modelling the shape of the distortions were approximated as
bar-shape profiles. Thus the defects were treated as rectangular potential
barriers as shown in Figure \ref{fig:Matrix}. 
The classical transport through the barrier states that the particle
may jump it over only if it has the energy greater than the potential
energy of the barrier. Or else the particle is reflected completely.
In quantum mechanics the probability of the particle to penetrate
through the finite size barrier is always greater than zero:
\begin{equation}
T=\left(1+\frac{V^{2}}{8E\left(V-E\right)}\left(e^{ka}-e^{-ka}\right)\right)^{-1}
\end{equation}

here $V$ is the potential height in energy units, $E$ is the particle
energy, $a$ is the width of the barrier in space units, $k$ is an
impulse parameter $k^{2}=2m\left(V-E\right)/\hbar^{2}$, relating
particle mass $m$ with energy parameters and fundamental constant.
The particle dispersion was limited to its ability to perform only
one constant space step and the fundamental constants were chosen
appropriate $\left(ka=\sqrt{V-E}\right)$. The reflection probability
is $R=1-T$. The infinite size of the barrier gives the same result
for the classical or quantum transport through the barrier. The modelling
was performed only for 2 dimensional space grid (size 1500 x 1500
points) as it is less time consuming than 3D but gives qualitatively
good results. The particle motion on the grid was available only in
4 directions. It was enough to have possibility for stepping over
all space points and every step may be treated as one dimensional,
this enabled simple evaluation of T for the next step. The step was
fixed to the length of the single space point. The defects were treated
either as neutral or screened to the size of the single space point.
Initially the space was randomly filled with the number of point defects
acting as background scatters. Further, more point defects were placed
in certain arrangements to create larger scale defects. The particle
for the modelling was randomly placed in the space at the grid point
without defect and counted $3\cdot10^{7}$ steps for each arrangement.
The translational symmetry boundary conditions were applied to keep
the particle in space. The potential of the single defect was assigned
to be 10 times grater than the energy of the jumping particle. So
the cluster potential may be several times greater than for the single
defect. In this case the particle energy corresponds to real transition
probability or low temperature of the system.

\section{Point Defects And Clusters}

\begin{figure}
\includegraphics{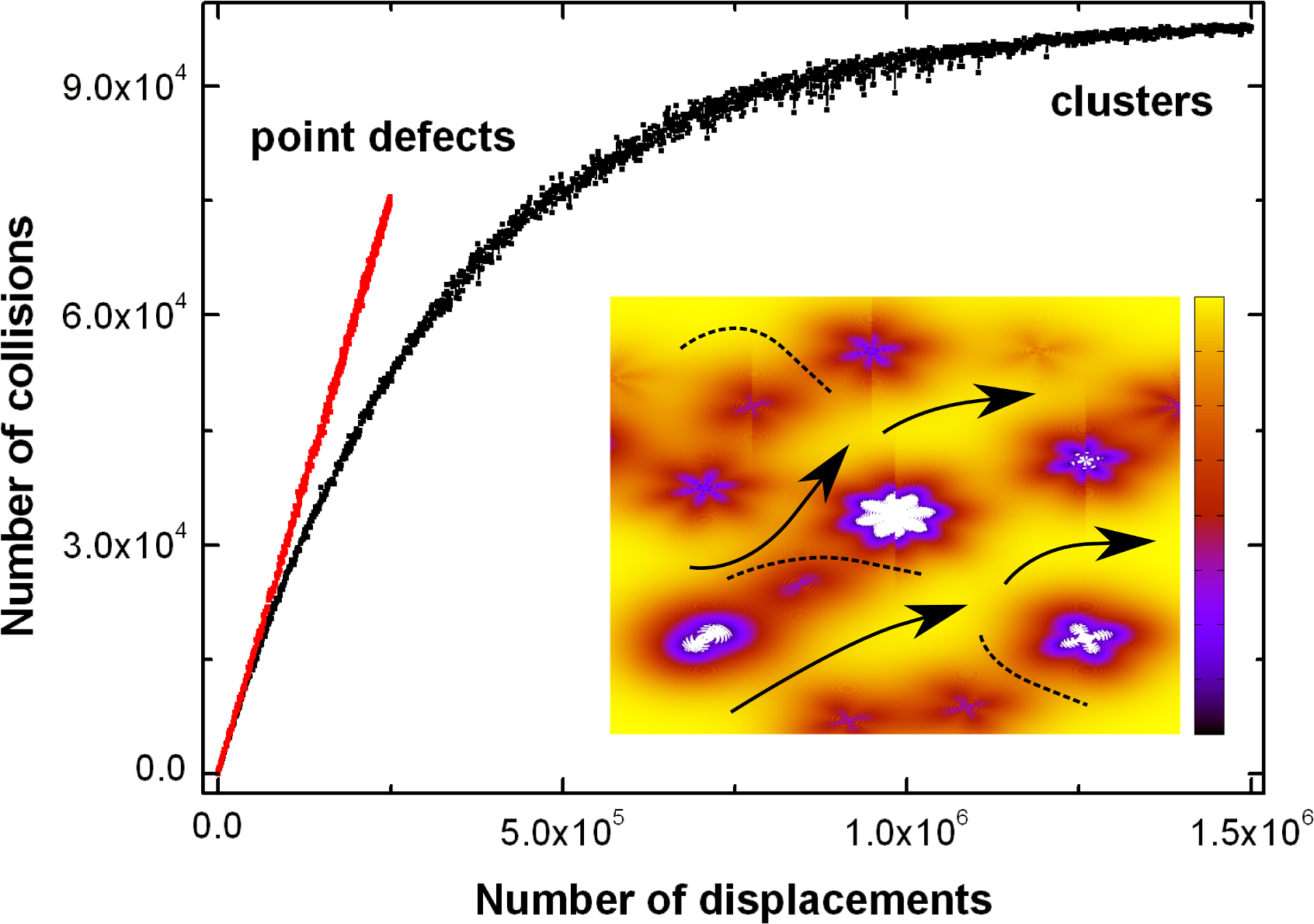}

\caption{Number of collisions in charge carrier transport dependence on the
number of point defects. The straight (red) dependency is related
to the single point defects in the modeling lattice. The curved (black)
dependency is related to the case when more than one point defect
is located in the same modeling lattice points. The inset illustrates
the paths of carrier transport around large conglomerations of defects.
Some paths are unavailable (dashed lines) because the field of defects
repel the carriers away, thus the defects screen the other defects.
(Color online)}

\label{fig:Clusters}
\end{figure}

When the crystal contains plenty of the point defects, their characterization
(detection) may be difficult because of their fields overlap and clusterization.
Starting with the model of randomly distributed point defects we may
observe that increasing the number of the scattering centres (point
defects) not linearly increases the number of the scattering events
as shown in Figure \ref{fig:Clusters}. 
Here the linear dependence corresponds to point defects. It was obtained
by putting only single defects at sites on the space grid. The defective
sites were allowed to appear in the neighbourhood, still their conglomerations
were not influencing as clusters. In contrast, the other dependency
is saturating, which indicates the presence of clusters. The illustration
in the Figure \ref{fig:Clusters} shows the paths of the charge carriers
around large scale defects (or clusters), which do not allow the carriers
to penetrate into some regions. Electrical characterization of such
material is complicated because the part of it is isolated while the
rest of it may still be of good quality. The electrical resistivity
of such material can be fairly measured but the carrier mobility measurements
may return failures. The clusters of only few point defects may be
considered as single scattering centres with larger interaction cross-section
because the electric field smoothers the edges of the branchy structures.

\section{Some Large Scale Deffect Geometries}

\begin{figure}
\includegraphics{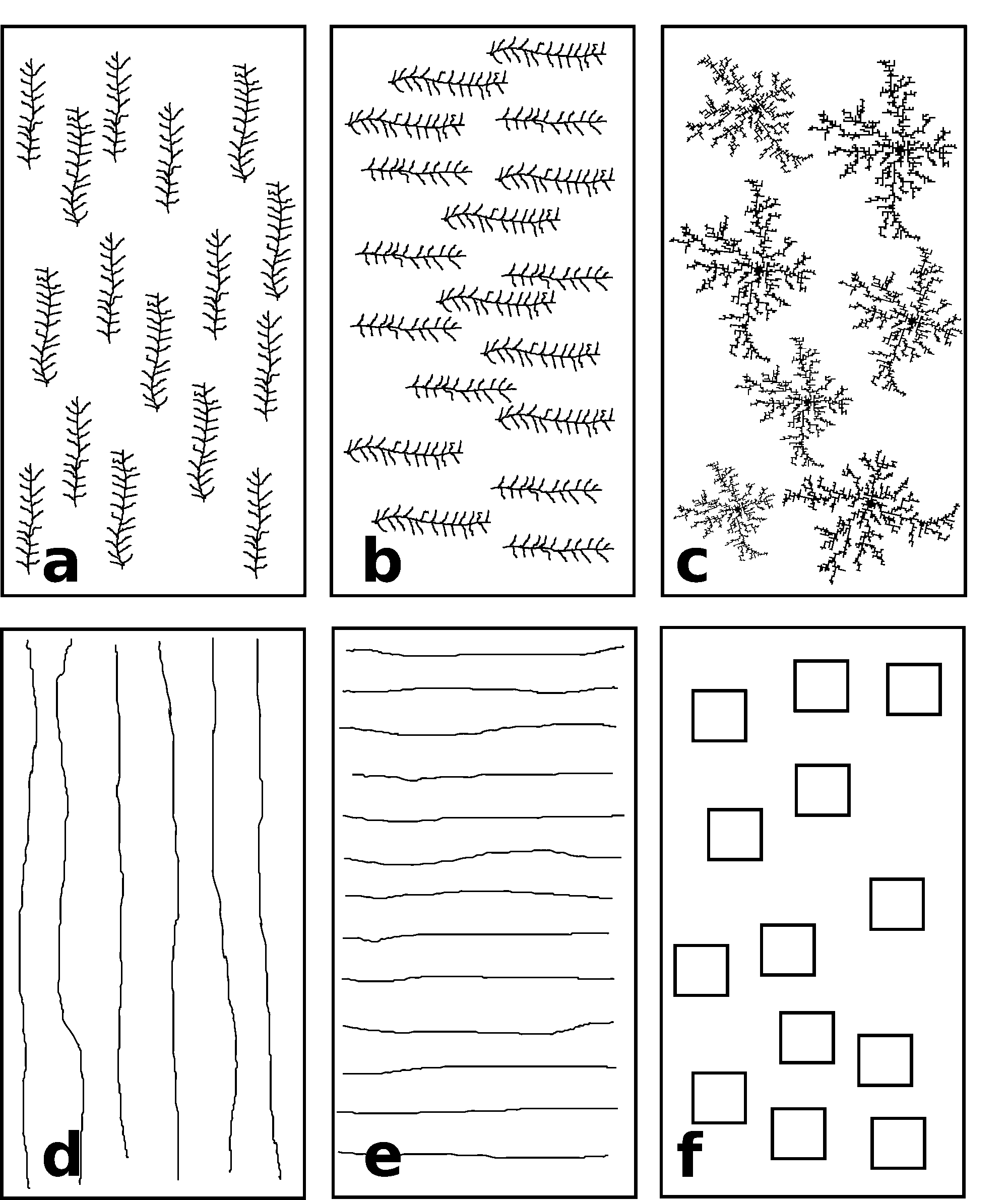}

\caption{Some large scale defect forms in the samples. The forms in parts (a)
and (b) correspond to twin boundaries in 3C-SiC \cite{key-3,key-4},
in (c) \textendash{} similar to \cite{key-1}, the ones in (d) and
(e) faced in \cite{key-2}, while (f) is artificial for comparison
to the others.}

\label{fig:SiCdef}
\end{figure}

\begin{figure}
\includegraphics{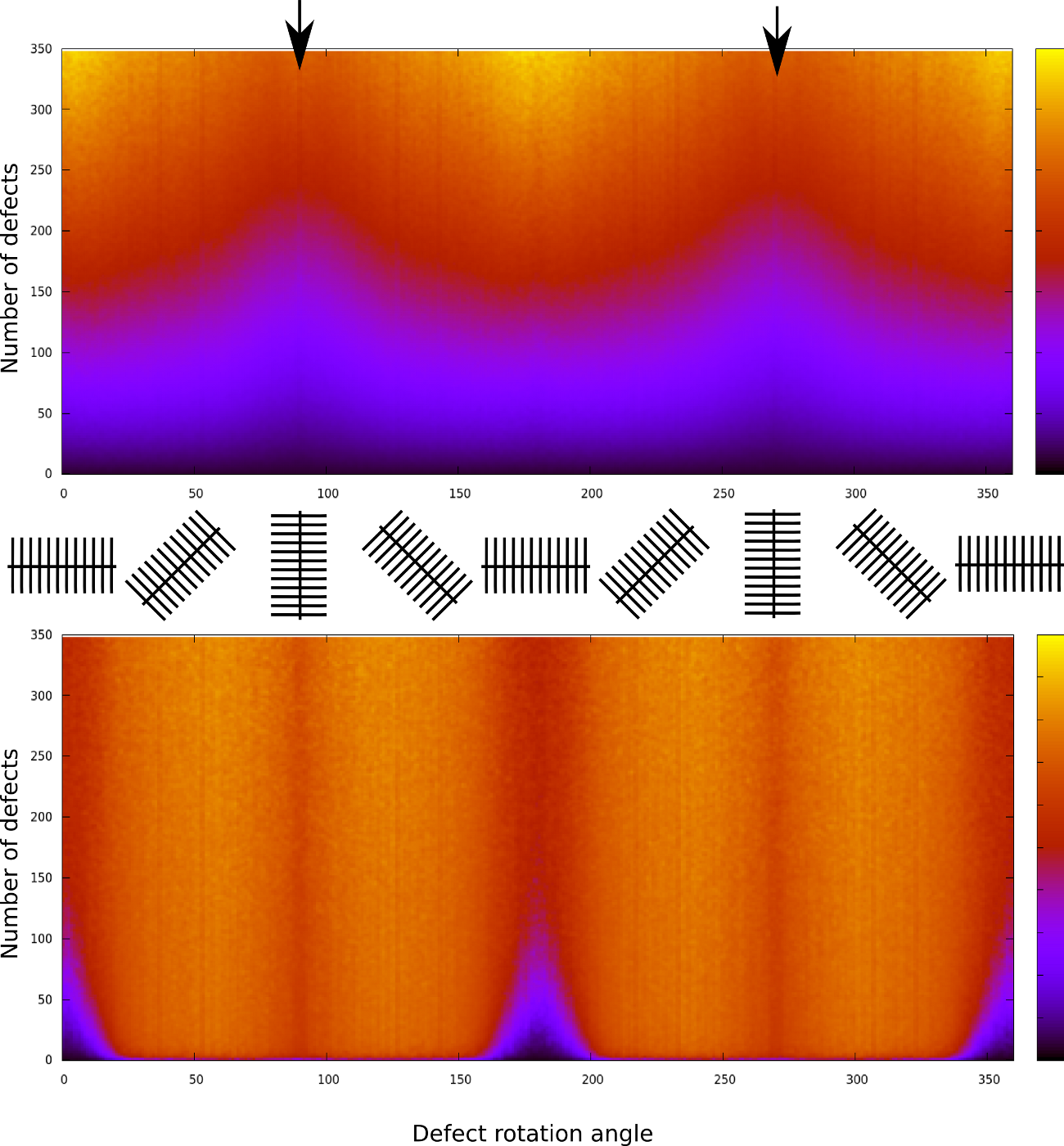}

\caption{The dependency of scattering events on defect number and their orientation.
The defect fir tree-like geometry and orientation is illustrated in
the middle. The upper plot is for the quantum transport regime and
the lower one is for the classical. The black arrows indicate the
minima\textquoteright{}s of the scattering events for the defect rotation
angles corresponding to the parallel orientation to the external field.
(Color online)}

\label{fig:rotate}
\end{figure}

Further geometries of this investigation are shown in Fig. 3. Here
the orientation of the defects (samples) to the external electric
voltage is important. The parts of the figures correspond to the crystal
samples with the electric potential driving carriers along the samples
length. The external voltage in the modeling is included into the
particle jump probability as additive parameter to the defect barrier
potential in one direction (x) and as additional jump direction choice
(x) for the 4 possible random directions (x, -x, y, -y) where the
barriers are absent. The parts (a) and (b) in Figure \ref{fig:SiCdef}
 sketch the pattern of defects obtained in the 3C-SiC crystals \cite{key-3,key-4}
(cubic lattice silicon carbide). These defects are the twin boundaries
(TBs) in a shape of fir tree branch. The magnetoresistant mobility
in these samples differs times depending on the orientation of the
samples. Experimentally only two orientations were measured as shown
in Figure \ref{fig:SiCdef}. Here the scattering modeling was applied
for these two orientations including the transformations from one
to another. The results are presented in Figure \ref{fig:rotate}.
The upper part of this figure corresponds to the quantum transport.
In the middle of the figure the orientation of the defects (sample)
is displayed. As one can notice, the zero angle (with the period of
180 degrees) gives the greatest scattering, while for the 90 degrees
(with the period) the scattering is lower. Depending on the number
of the large scale defects, the scattering difference on the rotation
angle becomes more distinct. The fir tree pattern of the defects acts
as carrier trap for the further scattering. In the quantum transport
regime if the particle once jumped over the barrier it is more probable
it will go away from the defect because of being dragged by the external
field. However in the classical regime, the particle may only walk
around the defect and the external field returns the particle back
into the defect of the complex structure. The lower part of the Figure
\ref{fig:rotate} 
shows clear difference between these transport regimes. The classical
one has the greatest scattering at the angles where the quantum one
has the lowest. The exception is marked by the arrows where the both
regimes have minima. This result of the different scattering behavior
is not trivial. In reality the barriers of the large scale defects
may be high enough to be treated as infinite (classical), but when
the defect \textquotedblleft{}captures\textquotedblright{} the carrier
its electric charge repulses other carriers away and statistically
the effective number of scatters changes. The described classical
transport may take place only when the carrier density is low and
the carriers do not interfere each other. Comparing the modeling results
with the experiment it may be stated that the TBs in 3C-SiC should
be treated quantum mechanically in the carrier transport analysis. 

\begin{figure}
\includegraphics{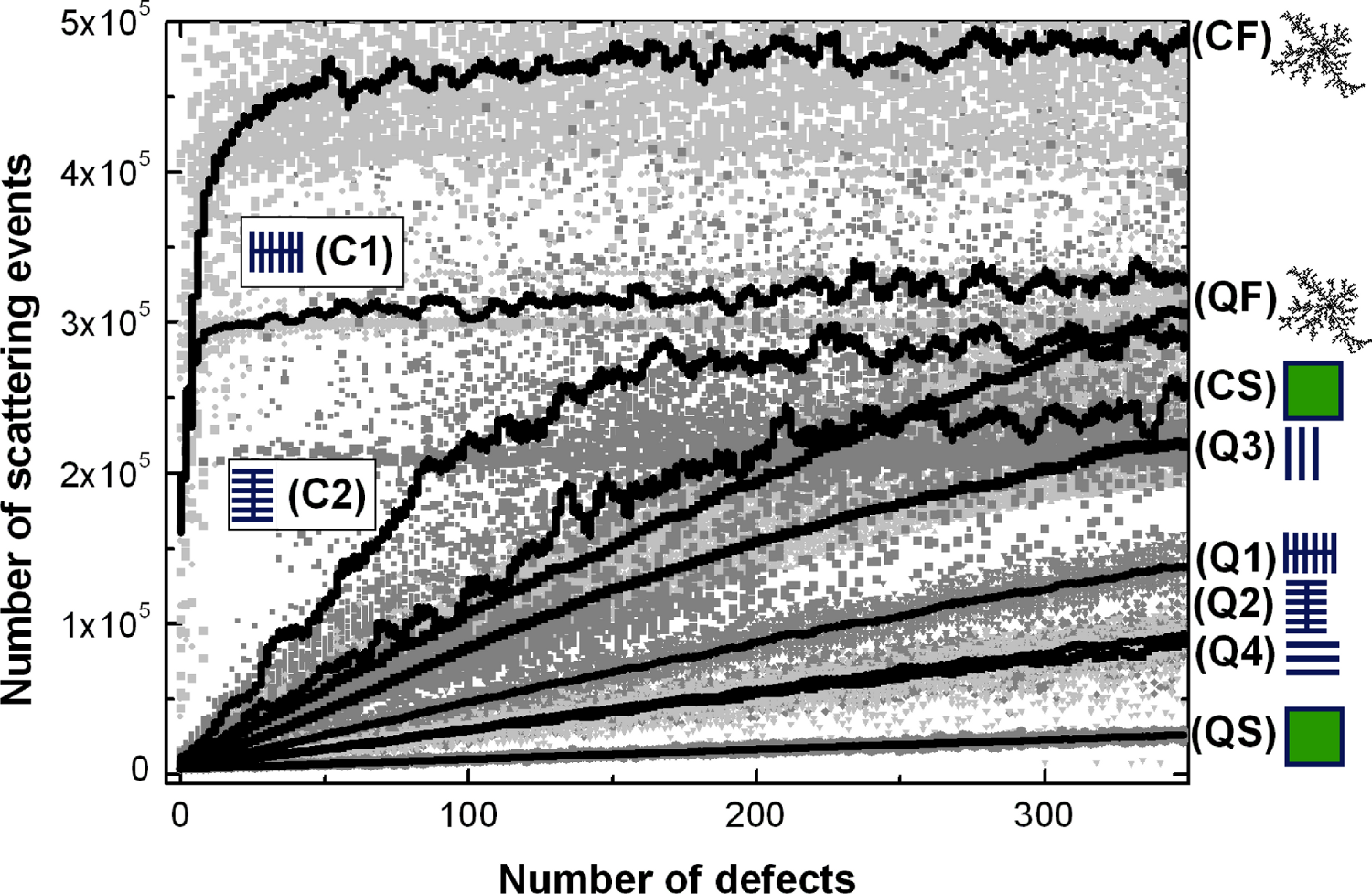}

\caption{The number of scattering events dependency on the number of large-scale
defects with shapes illustrated near the symbolic representations,
which mean the following: \textquotedblleft{}C\textquotedblright{}
\textendash{} classical, \textquotedblleft{}Q\textquotedblright{}
\textendash{} quantum transport model, \textquotedblleft{}F\textquotedblright{}
\textendash{} fractal, \textquotedblleft{}S\textquotedblright{} \textendash{}
square, \textquotedblleft{}1\textquotedblright{} the fir tree like
perpendicular to the external field, \textquotedblleft{}2\textquotedblright{}
\textendash{} the same but parallel to the field, \textquotedblleft{}3\textquotedblright{}
\textendash{} lines along the field, \textquotedblleft{}4\textquotedblright{}
\textendash{} lines perpendicular to the field. The solid curves (black)
are the averages of multiple events, which are displayed as single
points (grey tones). (Color online)}

\label{fig:statistic}
\end{figure}

The other experimentally measured sample with oriented defects is
shown in Figure \ref{fig:SiCdef} (d) and (e) parts \cite{key-2}.
These large scale defects were created by scratching the SiC crystal
substrate in lines then the 3C-SiC phase was grown on top and the
substrate removed. The boundaries between the 3C phase material parts
penetrated through the whole crystal. The idea of this scratching
was to release the strain of the overgrown layer and avoid cracks.
The good quality of the crystals between the inner boundaries was
obtained but the carrier transport through the whole crystal was greatly
modified. Here the good agreement of the experiment with the theory
was found. The defects along the sample contribute as more effective
scatters that the perpendicular ones. The modeling of only quantum
regime was analyzed because the perpendicular classical defects isolate
the material completely. The summary of these and some more defects
(from the Figure \ref{fig:SiCdef}) contribution for the carrier scattering
is shown in Figure \ref{fig:statistic}. 
At the right of the figure the defect forms are drawn and the names
are marked for quantum (with letter Q) and classical (C) regime. The
classical scattering is always greater comparing to the quantum one
for the same type of the defects. The growth of the number of the
scattering events grows nonlinearly with the number of defects indicating
the defect self screening effect. It is mostly expressed in the classical
transport, in which the number of the scattering events rapidly grows
to the saturation values. This case can be attributed to the material
with isolating inclusions, which do not influence scattering. By comparing
the types of the defects one can notice that the screening is most
significant in fractal type defects (Figure \ref{fig:SiCdef} c) or
fir tree-like defects perpendicular to the external field (Figure
\ref{fig:SiCdef} b). The similar to fractal defect structure may
be found in crystals with cracks or electrical break through. Also
the high energy irradiation of heavy particles may be followed by
cascade displacement events when point defect tracks are created,
which shape is close to shown in Figure \ref{fig:SiCdef}c. These
kind defects effectively screen the crystal volume from the charge
carriers like was shown in \cite{key-1}. The lowest contribution
to the scattering is found for the square shape defects in quantum
regime (Figure \ref{fig:statistic}, QS) and is similar to the form
described in \cite{key-6,key-7,key-8} for the Hall effect. In classical
regime their contribution is comparable to the other defect forms.
This is explained as the total screening of the interior of the large
defect in the classical regime. In case of the square defects (they
are empty inside) the quantum transport senses only their boundary
and background point defect scattering. Thus the number of active
point defects is lower comparing to the other defect shapes. The size
of each sort large scale defect was chosen to be of the similar space
volume. This fact does not restrict analysis to the single size of
the defects but makes them comparable between the sorts.

\section{Experimental Modeling}

\begin{figure}
\includegraphics{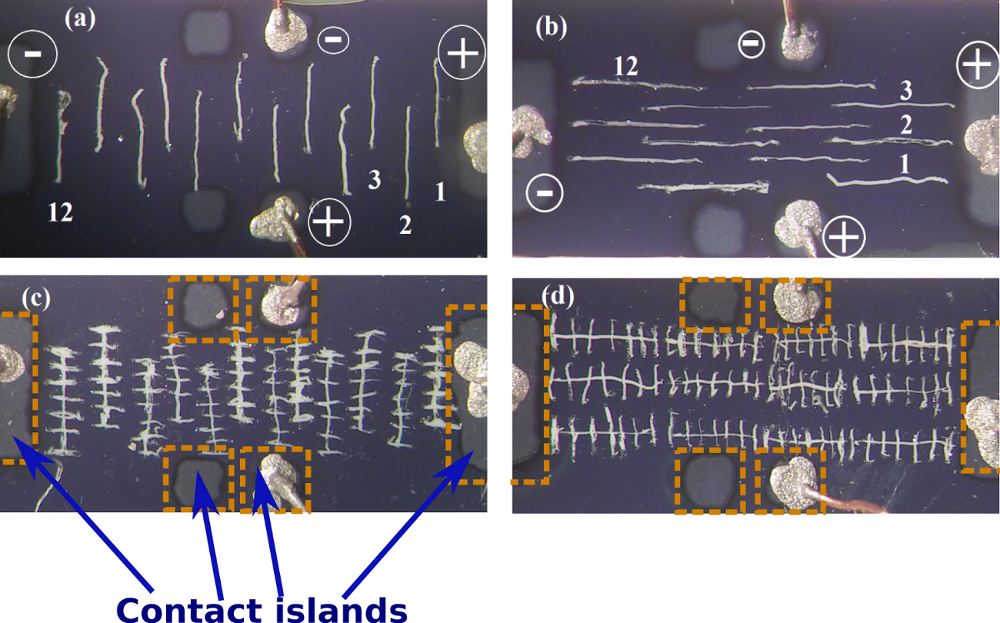}

\caption{Photographs of the thin semiconducting PbTe layers with electrical
contacts for Hall and magnetoresistivity measurements. The large-scale
defects are numbered in their manufacturing order is the same in all
samples. Polarity shows the power supply and voltmeter connection
arrangement. The distance between the left and the right contacts
is 5 mm. (Color online)}

\label{fig:foto}
\end{figure}

\begin{figure}
\includegraphics{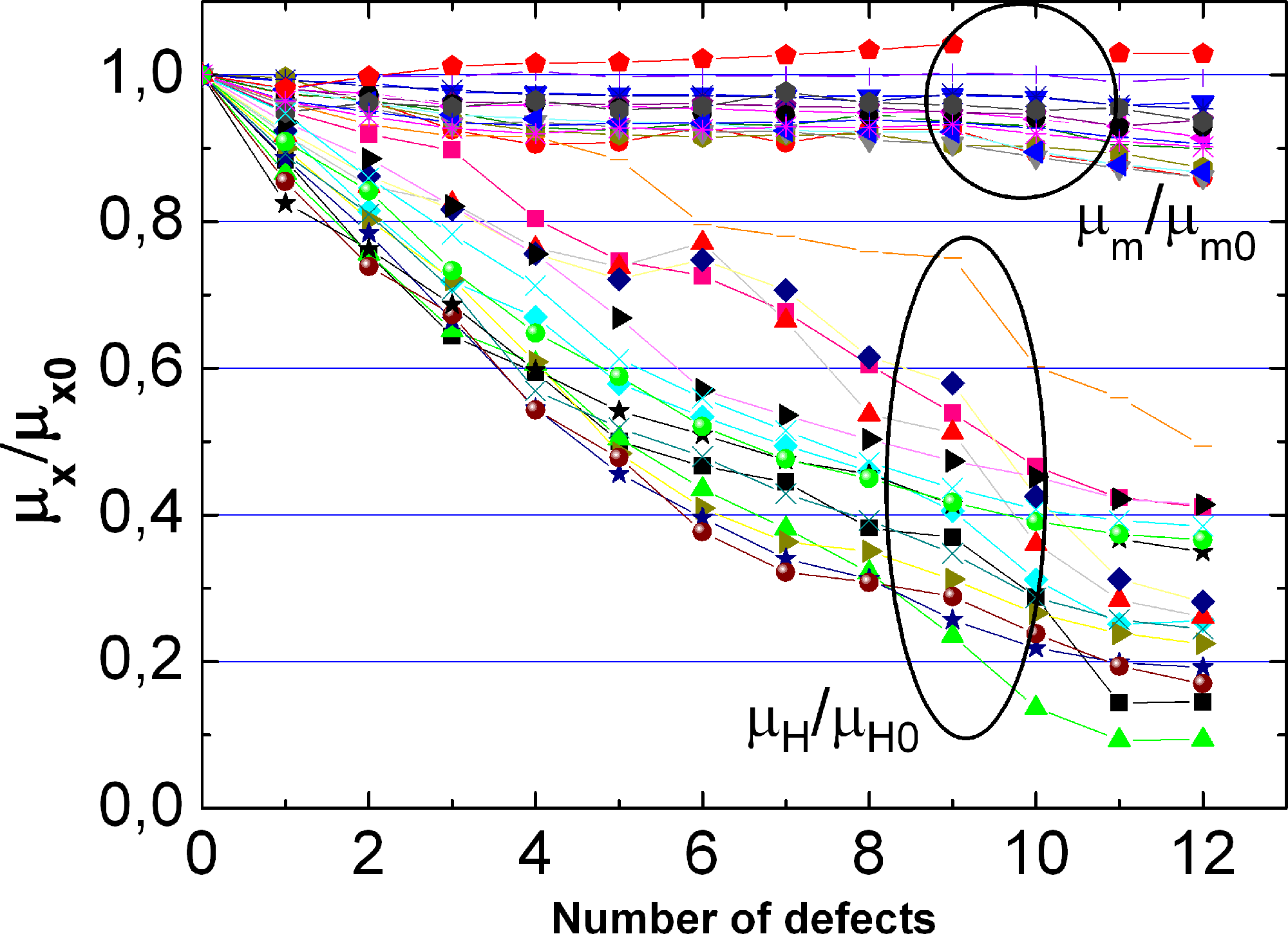}

\caption{Relative mobility dependence on the number of large scale defects
(all sorts and orientations). Upper curves calculated from magnetoresistivity
effect while the lower curves from Hall effect. (Color online) }

\label{fig:HallMR}
\end{figure}

\begin{figure}
\includegraphics{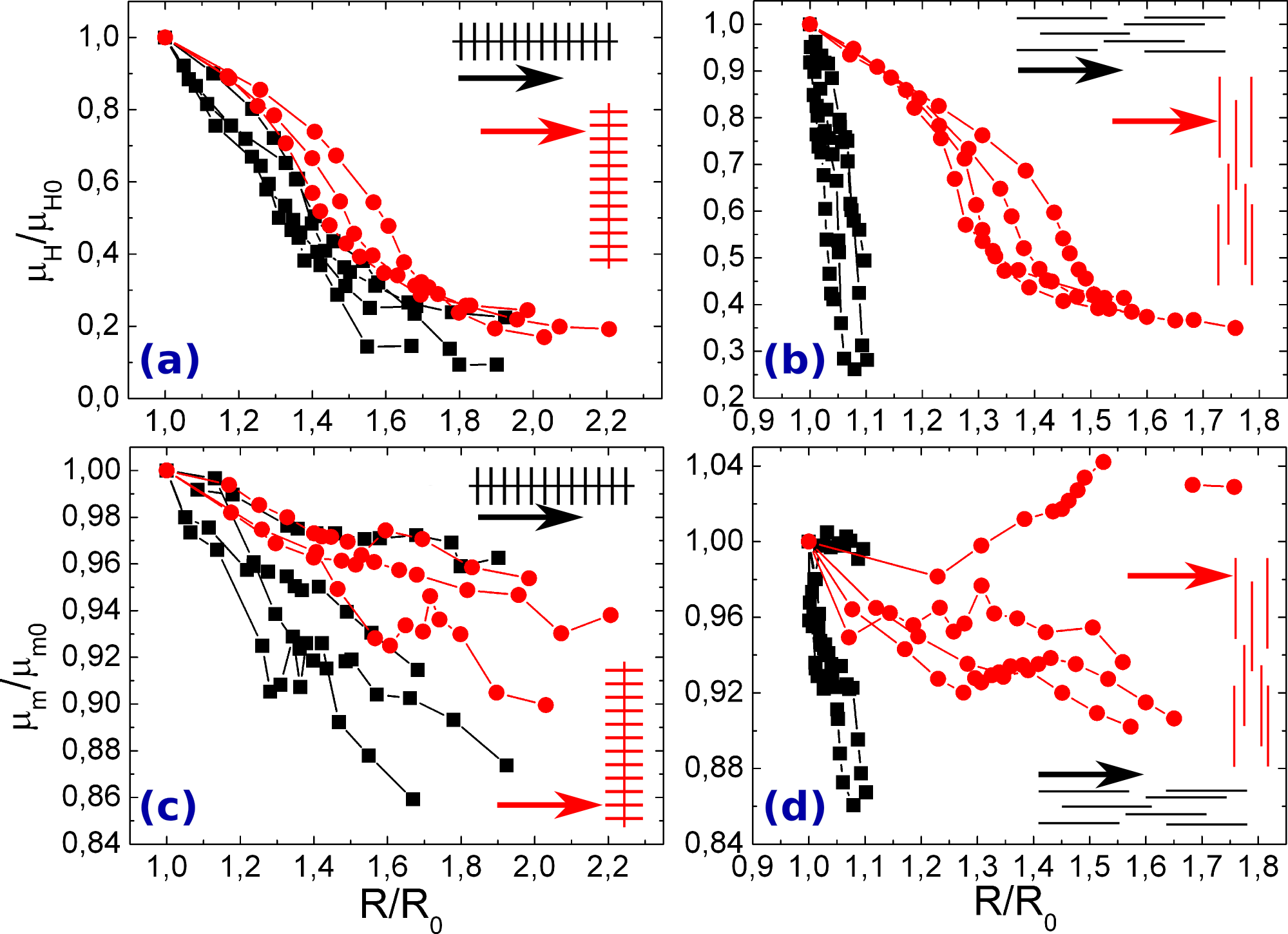}

\caption{Relative mobility dependence on relative resistivity. The large scale
defect orientations according to the current flows (arrows) are shown
for each case. Parts a and b correspond to Hall mobility, parts c
and d correspond to magnetoresistant mobility. Circles (red) points
correspond represent defect orientation perpendicular to the current
flow and squares (black) represent parallel orientation. (Color online)}

\label{fig:mobil4}
\end{figure}

To improve the analysis of the mentioned defects influence to the
charge transport parameters the experimental modeling was performed
on thin semiconducting layers. The layers produced by Physical Vapor
Deposition technique, which details are not important here. The semiconductor
material was chosen for convenience and ability was PbTe, which has
fair high carrier mobility and consequently produces higher Hall voltage.
In our experiment the layer formation conditions were such that the
highest Hall mobility was found about 750 $\textnormal{cm}^{2}/\textnormal{Vs}$.
For the defect modeling two types of the defects with two perpendicular
orientations were used as shown in Figure \ref{fig:foto}. 
Here four samples are shown with planar electrical contacts of Hall
bar-shape configuration and the scratched surfaces with intention
to repeat defect configuration from Figure \ref{fig:SiCdef} d, e,
a, b parts respectively. The scratches were done by metallic needle
in the numbered order shown on the photographs. After each large-scale
defect formation, Hall and magnetoresistivity measurements were performed
at room temperature to obtain carrier mobility. The experimental results
are shown in Figure \ref{fig:HallMR}. 
Here the relative mobility dependence on the number of scratched defects
is plotted for all samples and appear more like statistical variation.
The most distinguishing result is that Hall mobility signal (lower
curves in the figure grouped by an oval shape) is much more sensitive
to the appearance of the defects than magnetoresistant mobility (upper
curves). The similar result was observed in for the silicon irradiated
by neutrons \cite{key-1} or electrons \cite{key-9}. The difference
of magnetoresitant mobility values for the fir-tree like defects orientation
(shown in Figure \ref{fig:foto} c) and d) parts) was expected to
be the most expressed like in 3C-SiC case, however the experiment
showed that magnetoresistant mobility is not sensitive to this kind
defect orientation. The explanation may be that size of the structural
parts (perpendicular scratches) of the large scale defect does not
differ enough from the total size of the defect. In case of line-shape
defects (Figure \ref{fig:foto} a) and b) parts) the most interesting
observation is the increase of mangetoresistivity in certain sample
when defects are perpendicularly oriented to the average electric
current flow direction (Figure \ref{fig:foto}a). This result is in
agreement with the one found in 3C-SiC \cite{key-2}. The influence
of the defect shape and orientation in the current experiment may
be better observed for dependency on resistivity as shown in Figure
\ref{fig:mobil4}. 
The relation between resistivity and mobility may give information
about the charge carrier density. In this case that relation is not
straightforward and is affected by the geometry of the defect. One
can notice that in case of line-shape defects (Figure \ref{fig:mobil4}
b), the Hall mobility drops much faster with resistivity when the
defects are along the electric current and perpendicular to the Hall
field. The resistivity in this case cannot change much because the
defect lines separate the sample into array of parallel resistors.
However Hall voltage is created much smaller because the carriers
reach the Hall contacts through the field of obstacles. The same effect
is observed in magnetoresistivity (Figure \ref{fig:mobil4} d), when
line-shape defects do not allow carriers to bend in magnetic field
and increase their path. The opposite effect, when magnetoresitivity
is increased, is observed in the case when the same type defects are
oriented perpendicularly to the electric current path (Figure \ref{fig:mobil4}
d red dots). In this case magnetic field turns the carriers into the
regions around defects, which effectively increase the paths.

\begin{figure}

\includegraphics{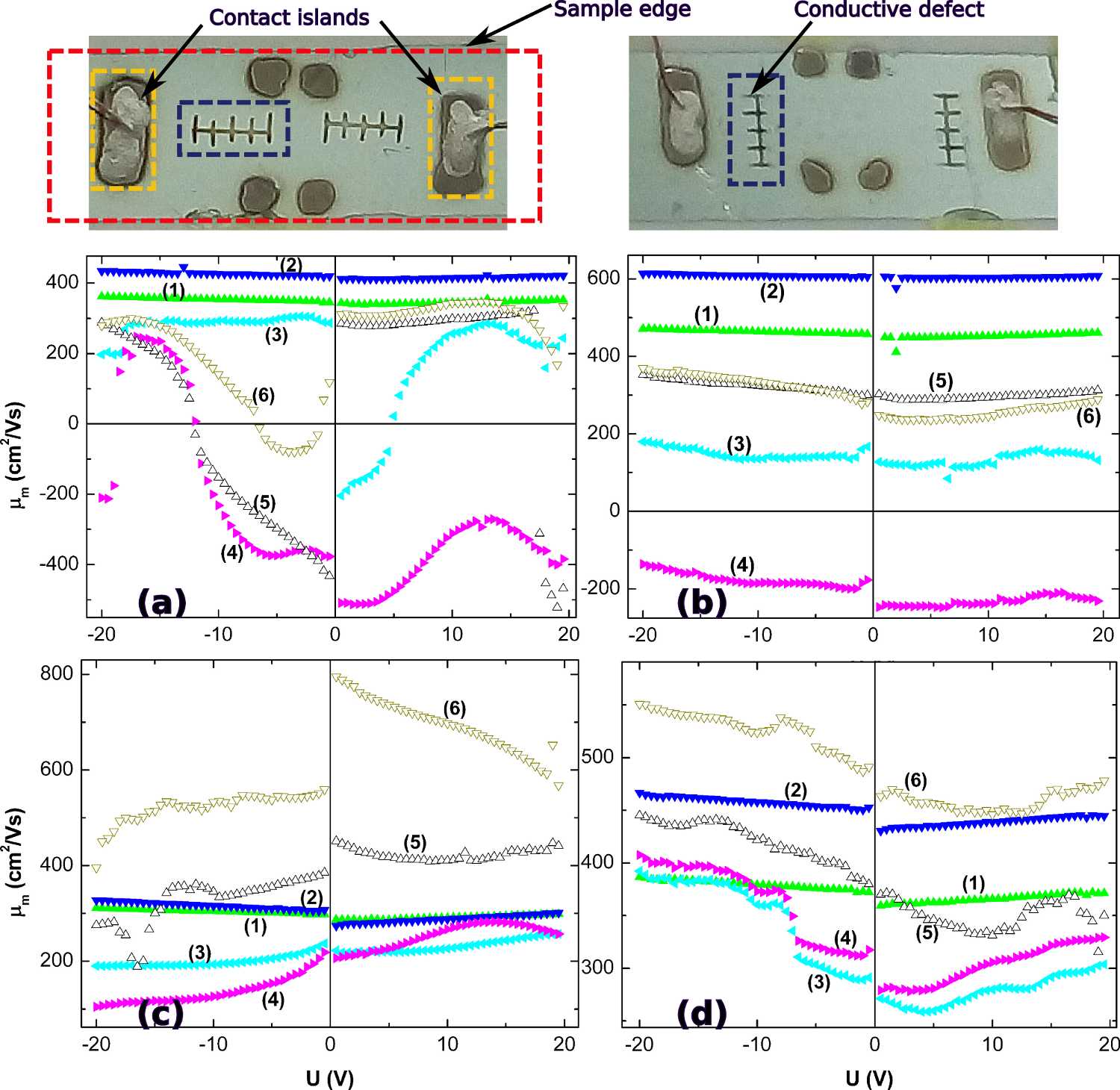}\caption{Magnetoresistant mobility dependence on applied voltage for the samples
with no artificial defects (1,2), one defect (3,4) and two defects
(5,6) in configuration shown at the top in the photographs. Parts
(a), (b), (c) and (d) correspond to different samples. The data plots
1,3,5 correspond to one magnetic field direction for the same sample
and the plots 2,4,6 correspond to the opposite magnetic field direction
in the same sample. (Color online) }
\label{fig:conductiveDef}

\end{figure}

The defects shown in Figure \ref{fig:foto} are electrical insulators
(at least in the bulk). It is reasonable to investigate the influence
of the same type defects to the electrical parameters when the defects
are electrical conductors. For this purpose the same material (PbTe)
was evaporated in vacuum with aluminium as shown in Figure \ref{fig:conductiveDef}
photographs. The size of the defects is of the order for consideration
about only electrical potential redistribution but not the scattering
of the charge carriers. In this case the orientation of the defect
is less important as one can nottice its influence to the magnetoresistivity.
In the figure the magnetoresistant mobility is calculated as the square
root of magnetoresistivity divided by magnetic field induction and
the sign extracted out from the root. When the defects are absent
the magnetoresistivity is proportional to the drift mobility with
certain geometry factor and it is independent on the applied electric
voltage (plots 1 and 2) as it should be. When only one defect is manufactured
the magnetoresistivity drops in all cases while even in some it becomes
negative (plots 3 and 4). The negative magnetoresistivity may be explained
by more conductive edges of the semiconductor layer when the carriers
are turned by magnetic field to these edges. It has to be noted that
the voltamperic characteristics are linear in all cases and the resistivity
is only slightly affected by the presence of the defect. Since the
artificial defect is formed from metal on semiconductor the junction
barrier should be present. The semiconductor layer below the metal
is not intacted, so the free carriers are influenced in these regions
only by potential redistribution, which may be considered as a bunch
of low barriers. In that point of view the scattering of the free
carriers may be also considered. The evaporation of the second defect
may be expected to enhance the described effect, however in all cases
it acts in the opposite manner: the magnetoresistivity is increased
and in some regions it is greater than the initial (for example part
c, positive voltage graphs 5 and 6 comparing to 1 and 2). It is similar
to the case for the defects shown in Figure \ref{fig:mobil4} part
(d) when the defects are perpendicular to the electric current direction.
This fact suggests that the increase of the magnetoresistivity is
more related to the defect which is repulsive (or isolating) to the
charge carriers. The influence of the defect orientation in this case
is observed in voltamperic characteristics. When the defect is as
shown in Figure \ref{fig:conductiveDef} at the left photograph, the
resistivity after the formation of the first defect increases while
the second defect slightly decreases the resistivity in comparison
to the previuous one. For the another orientation the resistivity
constantly increases with the number of defects. Since the doping
was constant in all cases the conductivity may change only with carrier
mobility, which is not proportional to the magnetoresistivity any
more.

\section{Summary}

The material characterization by mobility measurements reveals the
number of scattering centers, which may be not exact because the large-scale
defects may demonstrate self-screening effect. The complete and reliable
analysis requires a set of investigations with different numbers of
defects because their influence is not straight-forward for the extrapolation.
In some cases these numbers may be changed with intense light excitation,
when non-equilibrium charge carriers change the profiles of the defect
potentials. The shape and orientation of the large-scale defects may
play a significant role in carrier transport and the structural characterization
has to be performed before making conclusions from the electrical
measurements.

\subsection*{ACKNOWLEDGEMENTS}

The authors would like to thank prof. J. V. Vaitkus from Vilnius University
for suggestions and recommendations preparing this work. L. Andrulionis
is grateful to the Institute of Applied Research for support with
\textquotedblleft{}Viscakas\textquotedblright{} stipend. The support
from Research Council of Lithuanian is also acknowledged (Grand No.
MIP-15200).

\subsection*{}


\begin{thebibliography}{1}
\bibitem{key-1}J.Vaitkus, A.Mekys, J.Storasta. Analysis of microinhomogeneity
of irradiated Si by Hall and magnetoresistance effects. 10th RD50
Workshop on Radiation hard semiconductor devices for very high luminosity
colliders. Vilnius, 2007. 

\bibitem{key-2}P. ¦\v{c}ajev, A. Mekys, P. Malinovskis, J. Storasta,
M. Kato, and K. Jara¨i\={u}nas. Electrical parameters of bulk 3C-SiC
crystals determined by Hall effect, magnetoresistivity, and contactless
time-resolved optical techniques. Materials Science Forum Vols. 679-680
(2011) 157-160. doi:10.4028/www.scientific.net/MSF.679-680.157. 

\bibitem{key-3}R. Vasiliauskas, A. Mekys, P. Malinovskis, M. Syvajarvi,
J. Storasta, R. Yakimova, Influence of twin boundary orientation on
magnetoresistivity effect in free standing 3C\textendash{}SiC , Materials
Letters 74 (2012) 203. 

\bibitem{key-4}R Vasiliauskas, A Mekys, P Malinovskis, S Juillaguet,
M Syvajarvi, J Storasta and R Yakimova Impact of extended defects
on Hall and magnetoresistivity effects in cubic silicon carbide.,
J. Phys. D: Appl. Phys. 45 (2012) 225102. 

\bibitem{key-5}S. Data. Electronic Transport in Mesoscopic Systems.
Cambridge University Press (1997) 369. 

\bibitem{key-6}R. H. Bube. Interpretation Of Hall And Photo-Hall
Effects In Inhomogeneous Materials. Appl. Phys. Lett. 13 (1968) 136. 

\bibitem{key-7}J. Viscakas, K. Lipskis, A. Sakalas. On the interpretation
of Hall and thermoelectric effects in polycrystalline films. Lith.
Jour. Phys. 5 (1971) 799-806.

\bibitem{key-8}W.Siegel, S Schulte, G. Kühnel, J. Monecke. Hall mobility
lowering in undoped n-type bulk GaAs due to cellular-structure related
nonuniformities. J. Appl. Phys. 81 (1997) 3155. 

\bibitem{key-9}A. Mekys, V. Rumbauskas, J. Storasta, L. Makarenko,
J. V. Vaitkus. Defect analysis in fast electron irradiated silicon
by Hall and magnetoresistivity means., Nuclear Instruments and Methods
in Physics Research Section B 338 (2014) 95. \end{thebibliography}
\end{document}